\documentclass{article}
\usepackage{me}
\usepackage{wide}

\draftfalse

\newcommand{\cutval}[1]{{\cal C}(#1)}
\newcommand{\degreel}{\delta}
\newcommand{\degree}[1]{\degreel(#1)}
\newcommand{\degreesum}[1]{{\degreel^\downarrow(#1)}}
\newcommand{\crossweight}[2]{{\cal C}(#1,#2)}
\newcommand{\edgeweight}{\crossweight}

\newcommand{\treefix}[1]{#1^\downarrow}
\newcommand{\descendants}[1]{#1^\downarrow}
\newcommand{\ancestors}[1]{{#1}^\uparrow}

\newcommand{\bothbelowl}{\rho}
\newcommand{\bothbelow}[1]{{\bothbelowl^\downarrow(#1)}}
\newcommand{\lcaweight}[1]{{\bothbelowl(#1)}}

\newcommand{\cutvals}[2]{\cutval{\descendants{#1} \cup \descendants{#2}}}
\newcommand{\precut}[2]{{\cal C}_{#1}(#2)}
\newcommand{\minprecut}[1]{{\cal C}_{#1}}
\newcommand{\marker}[1]{{\tt val}[#1]}
\newcommand{\addpath}[2]{{\tt AddPath}(#1,#2)}
\newcommand{\minpath}[1]{{\tt MinPath}(#1)}

\newcommand{\incomparable}[2]{{#1 \perp #2}}
\newcommand{\treedeg}{D}
\newcommand{\cutdeg}{d}

  \title{Minimum Cuts in Near-Linear Time}

  \author{David R. Karger\thanks{\sloppy MIT Laboratory for Computer Science,
      Cambridge, MA 02138.   \protect\newline email: 
      {\tt karger\atsign{}lcs.mit.edu} \protect\newline URL: {\tt
      http://theory.lcs.mit.edu/$\sim$karger}.  \protect\newline
      Research supported in 
      part by ARPA contract  N00014-95-1-1246 and NSF contract CCR-9624239.}}

\begin{document}

  \maketitle

  \begin{abstract}
    We significantly improve known time bounds for solving the
    minimum cut problem on undirected graphs.  We use a ``semi-duality''
    between minimum cuts and maximum spanning tree packings combined
    with our previously developed random sampling techniques.  We give
    a randomized algorithm that finds a minimum cut in an $m$-edge,
    $n$-vertex graph with high probability in $O(m\log^3 n)$ time.  We
    also give a simpler randomized algorithm that finds {\em all}
    minimum cuts with high probability in $O(n^2\log n)$ time.  This
    variant has an optimal $\RNC$ parallelization.  Both variants
    improve on the previous best time bound of $O(n^2\log^3 n)$.
    Other applications of the tree-packing approach are new, nearly
    tight bounds on the number of {\em near minimum} cuts a graph may
    have and a new data structure for representing them in a
    space-efficient manner.
  \end{abstract}

\section{Introduction}
\label{sec:intro}

The minimum cut problem has been studied for many years as a
fundamental graph optimization problem with numerous applications.
Initially, the problem was considered a harder variant of the $s$-$t$
minimum cut problem and was solved in $\Olog(mn^2)$
time ($\Olog(f)$ denotes $O(f\polylog f)$) on $m$-edge,
$n$-vertex graphs~\cite{Ford:Flowbook,Gomory:FlowTree}.  Improvements
then followed: first to
$\Olog(mn)$ time~\cite{Hao:Mincut,Nagamochi:Mincut}, showing
that minimum cuts were as easy to find as maximum flows; then to
$\Olog(n^2)$ time~\cite{Karger:Contraction}, showing
they were significantly easier.

In this paper we give an algorithm with an $O(m\log^3 n)$ running
time.  Initially perceived as a harder variant of the maximum flow
problem, the minimum cut problem turns out to be solvable in
near-linear time.  Side effects of our analysis include new
combinatorial theorems on the structure, enumeration and
representation of minimum cuts.  We also give a relatively simple
algorithm that runs in linear time on a large class of graphs and runs
in $O(n^2\log n)$ time on all graphs.  This algorithm still dominates
all previous ones and can also be parallelized optimally (that is,
with time-processor product no worse than the sequential algorithm's)
to yield the best-known $\RNC$ algorithm for the problem.

Our algorithm is based on a ``semi-duality'' between minimum cuts and
undirected spanning tree packings.  It is therefore fundamentally
different from past approaches based on flows or on edge contraction.
An algorithm developed by Gabow~\cite{Gabow:Connectivity} uses {\em
  directed} tree packing to solve the problem, but ours appears to be
the first to use {\em undirected} spanning tree packings.  It thus
introduces a different approach to finding minimum cuts in undirected
graphs.

\subsection{The Problem}

This paper studies the {\em minimum cut} problem.  Given a graph with
$n$ vertices and $m$ (possibly weighted) edges, we wish to partition
the vertices into two non-empty sets so as to minimize the number (or
total weight) of edges crossing between them.  More formally, a {\em
  cut} $(A,B)$ of a graph $G$ is a partition of the vertices of $G$
into two nonempty sets $A$ and $B$.  An edge $(v,w)$ {\em crosses} cut
$(A,B)$ if one of $v$ and $w$ is in $A$ and the other in $B$.  The
{\em value} of a cut is the number of edges that cross the cut or, in
a weighted graph, the sum of the weights of the edges that cross the
cut.  The minimum cut problem is to find a cut of minimum value.  We
use $c$ to denote the value of this minimum cut.  Throughout this
paper, the graph is assumed to be connected, since otherwise the
problem is trivial.  We also require that the edge weights be
non-negative, because otherwise the problem is $\NP$-complete by a
transformation from the maximum-cut problem~\cite[page 210]{Garey:NP}.
We distinguish the minimum cut problem from the {\em $s$-$t$ minimum
  cut problem} in which we require that two specified vertices $s$ and
$t$ be on opposite sides of the cut; in the  minimum cut problem
there is no such restriction.

\comment{
Application areas that use minimum cuts include information
retrieval~\cite{Botafogo:Hypertext}, network reliability
analysis~\cite{Karger:Reliability}, and cutting plane algorithms for
combinatorial optimization problems such as the Traveling Salesman
Problem~\cite{Dantzig:TSP,Lawler:TSP,Padberg:Mincut,Applegate:Private}
and others~\cite{Picard:Mincut}.
}

\subsection{Applications}

The minimum cut problem has several applications.  Picard and
Queyranne~\cite{Picard:Mincut} survey applications including graph
partitioning problems, the study of project networks, and partitioning
items in a database.  In information retrieval, minimum cuts have been
used to identify clusters of topically related documents in hypertext
systems~\cite{Botafogo:Hypertext}.  The problem of determining the
connectivity of a network arises frequently in issues of network
design and network reliability (one of our previous
papers~\cite{Karger:Reliability} exploits an extremely tight
connection between minimum cuts and network reliability).  Minimum
cut computations are used to find the {\em subtour elimination
  constraints} that are needed in the implementation of cutting plane
algorithms for solving the traveling salesman
problem~\cite{Dantzig:TSP,Lawler:TSP}.  Padberg and
Rinaldi~\cite{Padberg:Mincut} and Applegate~\cite{Applegate:Private}
have reported that solving min-cut problems was the
computational bottleneck in their state-of-the-art cutting-plane based
TSP algorithm, as well as other cutting-plane based algorithms for
combinatorial problems whose solutions induce connected graphs.

\subsection{Past Work}

The minimum cut problem was originally approached as a variant of the
$s$-$t$ minimum cut problem and solved using flow
techniques~\cite{Ford:Flowbook}.  The most obvious method is to
compute $s$-$t$ min-cuts for all vertices $s$ and $t$; this requires
$\binom{n}{2}$ flow computations.  Gomory and
Hu~\cite{Gomory:FlowTree} showed that the problem could be solved by
$n$ flow computations; using present flow
algorithms~\cite{Goldberg:Maxflow} this gives a time bound of
$\Olog(mn^2)$.  After a lengthy period of little progress, several new
algorithms appeared.  Hao and Orlin~\cite{Hao:Mincut} showed how the
$n$ flow computations could be performed in the time for one,
improving the running time to $\Olog(mn)$.
Gabow~\cite{Gabow:Connectivity} gave an alternative
``augmenting-tree'' scheme that required $\Olog(mc)$ time on an
unweighted graph with minimum cut $c$.

Sampling as an attack on minimum cuts was introduced
in~\cite{Karger:Skeleton}.  We showed that sampling could be used
to find approximate minimum cuts in $\Olog(m)$ time and to find them
exactly in $\Olog(m\sqrt{c})$ time (combining sampling with Gabow's
algorithm).  The sampling technique is crucial to this paper as well.

The above approaches all use maximum flow techniques that treat
undirected graphs as directed graphs (Gabow's algorithm packs directed
trees rather than directed paths, but still addresses directed
graphs).  Recently a flowless undirected-graph approach based on {\em
  edge contraction} was discovered.  If we can identify an edge that
does not cross the minimum cut, then we can merge its endpoints into a
single new vertex without affecting the minimum cut.  Performing $n-2$
valid contractions will reduce the graph to two vertices, at which
point there is only one nontrivial cut that must therefore correspond
to the minimum cut.  Nagamochi and
Ibaraki~\cite{Nagamochi:Connectivity,Nagamochi:Mincut} devised a {\em
  sparse certificate} computation that found a contractible edge in
$O(m)$ time; this led to an $O(mn)$-time minimum cut algorithm.
Karger and Stein~\cite{Karger:Contraction} showed that random edge
contraction worked well, leading to an algorithm running in $O(n^2 \log^3 n)$
time.

\subsection{Our Results}

In this paper we present a different approach to minimum cuts that
yields faster algorithms than were previously known. We use a
``semi-duality'' between minimum cuts and maximum packings of
undirected spanning trees---arguably a more natural dual than the
directed flows or directed spanning trees used
previously~\cite{Gomory:FlowTree,Hao:Mincut,Gabow:Connectivity}. Our
approach does not rely fundamentally on flows {\em or} edge
contractions. For a weighted, undirected graph we give:
\begin{itemize}
\item A randomized algorithm that finds a minimum cut with constant
  probability in $O(m\log^2 n)$ time and with high probability in
  $O(m\log^3 n)$ time.  This significantly improves the previous best
  $O(n^2\log^3 n)$ bound of~\cite{Karger:Contraction}.  A technical refinement
  improves the time bound by an additional $\log\log n$ factor.
\item A variant that finds a minimum cut with constant probability in
  $O(n^2)$ time and finds {\em all} (possibly $\Omega(n^2)$) minimum
  cuts with high probability in $O(n^2\log n)$ time.  While this only
  gives a speedup of $\Theta(\log^2 n)$ over the algorithm
  of~\cite{Karger:Contraction}, the new algorithm is quite simple,
  relying only on computing minimum spanning trees, finding least
  common ancestors and evaluating expression-trees.  It runs in $O(m)$
  time on a natural class of graphs.
\item A parallel version of our simpler algorithm that runs in
  $O(\log^3 n)$ time using $n^2/\log^2 n$ processors.
\item New bounds of~$\Theta(n^{\lfloor 2\alpha \rfloor})$ on the
  number of cuts with value at most $\alpha$ times the minimum that a
  graph can have.  For non-half-integral $\alpha$ this improves on a
  previous best general bound of
  $O(n^{2\alpha})$~\cite{Karger:Mincut}.  This bound is the first to
  match the ``step function'' behavior of the known lower bounds with
  respect to $\alpha$.
\item A new data structure that represents all minimum and
  near-minimum cuts.  Besides being smaller than previous
  representations~\cite{Karger:Contraction}, this representation
  highlights certain structural features of a graph's near-minimum
  cuts.
\end{itemize}

\subsection{Our Methods}

We now summarize the approach of the paper.  At its core is the
following definition:

\begin{definition}
  Let $T$ be a spanning tree of $G$.  We say that a cut in $G$ {\em
    $k$-respects} $T$ if it cuts at most $k$ edges of $T$.  We also
  say that $T$ {\em $k$-constrains} the cut in $G$.
\end{definition}

Nash-Williams~\cite{Nash:Trees} proved that any any graph with minimum
cut $c$ contains a set of $c/2$ edge-disjoint spanning trees.  These
trees must divide up the $c$ minimum cut edges.  It follows that any
such {\em tree packing} contains a tree that 2-constrains the minimum
cut.  Using this idea, we show that for any graph $G$, we can find (by
packing trees) a small set of spanning trees such that every minimum
cut $2$-respects some of the spanning trees.  This lets us reduce to
the following problem: find the minimum cut in $G$ that $2$-respects a
given spanning tree.  While this might appear a harder problem than
the original, it turns out that the added constraints make the problem
easier.  Our scheme can be thought of as a variant of the ``branch and
bound'' techniques used for hard optimization problems: we consider
several families of added constraints, one of which is satisfied by
the optimum solution.  We can find the optimum by solving each
constrained optimization problem.

To introduce our approach, we show in Section~\ref{sec:Nash-Williams}
that a maximum packing of spanning trees always contains many trees
that $2$-constrain any minimum cut.  An immediate application of this
observation in Section~\ref{sec:combinatorics} yields the tightest
known bounds on the number of near-minimum cuts in a graph.

Next we turn to the problem of actually finding the minimum cut.  It
follows from the previous paragraph that we can find the minimum cut
by examining each tree of a maximum packing and finding the smallest
cut that $2$-respects each tree.  Although finding a maximum tree
packing is hard, we show in Section~\ref{sec:packing} that Gabow's
minimum cut algorithm~\cite{Gabow:Connectivity} or the
Plotkin-Shmoys-Tardos fractional packing
algorithm~\cite{Plotkin:Packing} can be used to find a packing that is
sufficient for our purposes.  Unfortunately, in a graph with minimum
cut $c$, such a packing contains roughly $c$ trees, so checking all of
them is prohibitively slow.  Indeed, even finding the packing (with either
algorithm) takes time proportional to $c$.  To eliminate this factor,
we use random sampling techniques that we developed
previously~\cite{Karger:Skeleton,Karger:Thesis} to reduce the packing
time to $O(m+n\log^3 n)$ and the number of trees in the packing to
$O(\log n)$.

Once we have the packing, it remains to examine each tree in it.  We
begin in Section~\ref{sec:tree 1 slow} by showing how to find the
smallest cut that {\em 1-respects} a given tree in $O(m)$ time.  Our
algorithm involves a simple dynamic program that determines the value
of the cut ``induced'' by removing each edge of the given tree.
Besides introducing our techniques, this algorithm can be used to find
the minimum cut in $O(m+n\log^3 n)$ time on a natural family of
``fat'' graphs containing many spanning trees.

In some cases we will not be able to find a tree that $1$-constrains
the minimum cut, so we move on to finding the minimum cut
$2$-respecting a given tree.  We present two solutions.  In
Section~\ref{sec:tree 2 slow} we extend the simple dynamic program of
Section~\ref{sec:tree 1 slow} to compute the value of the cut induced
by removing each {\em pair} of tree edges.  The algorithm runs in
$\Theta(n^2)$ time---optimal for an algorithm that computes the values
for all $\binom{n }{2}$ pairs.  By running on $O(\log n)$ trees it
identifies {\em all} minimum cuts with high probability in $O(n^2\log n)$
time.  In Section~\ref{sec:parallel}, we also show how it can be
parallelized optimally.

The second algorithm, presented in Section~\ref{sec:tree 2 fast},
implicitly considers all $\binom{n}{2}$ pairs of tree edges without
taking $\Omega(n^2)$ time to do so.  We show how dynamic tree data
structures~\cite{Sleator:DynamicTrees} can be used to evaluate a
particular tree in $O(m\log^2 n)$ time.  Thus evaluating the necessary
$O(\log n)$ trees takes $O(m\log^3 n)$ time and gives the claimed bound.

We finish by giving some extensions of the algorithm: a fast and
simple algorithm for a large class of ``fat'' graphs, and some
refinements that improve the running time of our complicated algorithm
by some small factors.

The initial tree-packing step is the only one using randomization; all
subsequent operations in the algorithm are deterministic.  The fact
that random sampling is used means that our algorithms, like those
of~\cite{Karger:Contraction}, are {\em Monte Carlo} algorithms.
Although the probability of success is high, there is no fast way to
certify the correctness of the final answer (indeed, the fastest known
method is to use one of the deterministic minimum cut algorithms).

\section{Cuts and Tree Packings}
\label{sec:Nash-Williams}

In this section, we give some background on tree packings and
their connection to cut problems.  For the moment, let us restrict
attention to unweighted graphs.  An unweighted {\em tree packing} is a
set of edge-disjoint spanning trees in a graph.  The {\em value} of
the packing is the number of trees in it.  Clearly, every spanning
tree in an unweighted packing must use at least one edge of every cut.
Thus, at most $c$ trees can be packed in a graph with minimum cut
$c$.  Nash-Williams~\cite{Nash:Trees} gave a related lower bound:

\begin{theorem}[\cite{Nash:Trees}]
\labelprop{Theorem}{prop:trees}
Any undirected graph with minimum cut $c$ contains a tree packing of
value at least $c/2$.
\end{theorem}

Nash-Williams actually proved a tighter result: if $c_r$ is the value
of the minimum $r$-way cut in the graph, then the value of the maximum
tree packing is exactly $\floor{\min_r c_r/(r-1)}$.  But observe that
the minimum $r$-way cut must have $c_r \ge rc/2$, since each of the
$r$ components of the cut must have $c$ edges leaving it.  Thus
\[
\min \frac{c_r}{r-1} \ge \min \frac{rc}{2(r-1)} \ge c/2
\]
\ref{prop:trees} is not universally tight---for example, a single tree
has minimum cut 1 and contains one spanning tree.  However, it is
existentially tight since, for example, a cycle has minimum cut $2$
but has maximum tree packing value 1.  We will find later that the
tightness of a graph with respect to tree packings is a factor that
determines the hardness of finding its minimum cuts using our
algorithm.  The factor-of-2 gap is particular to undirected graphs:
Edmonds~\cite{Edmonds:Branchings} proved that a {\em directed} graph
with {\em directed} minimum cut $c$ has a packing of exactly $c$
edge-disjoint directed spanning trees (Gabow exploits this in his
minimum cut algorithm~\cite{Gabow:Connectivity}).

Since any graph has a packing with $c/2$ trees, it follows that a
maximum packing will contain at least $c/2$ trees.  Consider any
minimum cut.  Since the edges of the minimum cut must be partitioned
among the trees of the maximum packing, the average number of min-cut
edges per tree is at most $c/(c/2)=2$.  It follows that at least one
of the trees in such a packing has at most $2$ minimum-cut edges.  In
other words at least one tree $2$-constrains the minimum cut. 

We will strengthen this argument in \ref{prop:packing}.  First
let us see why it is useful.  If we are given a tree, and are able to
determine which of its edges cross the minimum cut, then in fact we
will have found the minimum cut.  Consider a tree in which all minimum
cut edges have been marked.  Start at any vertex and traverse the tree
path to any other vertex.  Each time we cross a min-cut edge, we know
that we have switched sides of the minimum cut.  It follows that two
vertices are on the same side of the minimum cut if and only if the
number of minimum cut edges on the path between them is even.  More
concretely, suppose a tree has only one minimum cut edge.  Then
removing this edge separates the tree into two subtrees, each of which
is one side of the minimum cut.  If a tree contains two minimum cut
edges, then removing them separates the tree into three pieces.  The
``central'' piece that is adjacent to both other pieces forms one side
of the minimum cut, while the two remaining pieces form the other
side.

The above paragraph demonstrates that any subset of the tree
edges defines a unique two-way cut of the graph $G$.  It is also clear
that every cut defines a unique set of tree edges---namely, those
crossing the cut---and that this correspondence between cuts and sets
of tree edges is a bijection.

\subsection{Weighted Packings}

We now extend the above discussion to argue that {\em many} of the
trees in a maximum packing must 2-respect that minimum cut.  This is
useful since it means that a randomly selected tree is likely to
2-constrain the minimum cut.  We also\marnote{later!} generalize
the tree packing definition to allow for weighted trees packed into
weighted graphs.\marnote{capacity for edge weights?}

\begin{definition}
  A (weighted) {\em tree packing} is a set of spanning trees, each
  with an assigned weight, such that the total weight of trees
  containing a given edge is no greater than the weight of that edge.
  The {\em value} of the packing is the total weight of the trees in
  it.
\end{definition}

If all edge and tree weights are integers, we can treat a weight-$w$
edge as a set of $w$ parallel edges, and a weight $w$ tree as a set of
$w$ identical trees.  This recovers the unweighted notion of a tree
packing as a set of edge-disjoint trees.  The value of the packing
becomes the number of (unweighted) trees.  The definition also allows
for fractional weights.  This can be ignored, though, since we can
always multiply all edge and tree weights by the least common
denominator of the fractions to return to the integral version.  It
follows that Nash-Williams' \ref{prop:trees} applies unchanged to
weighted tree packings.

\begin{lemma}
  \labelprop{Lemma}{prop:packing} Given any weighted tree packing of
  value $\beta c$ and any cut of value $\alpha c$, at least a
  $\frac12(3-\alpha/\beta)$ fraction of the trees (by weight)
  $2$-constrain the cut.
\end{lemma}
\begin{corollary}
In any maximum packing, half the trees (by weight) 2-constrain the
minimum cut.
\end{corollary}
\begin{proof}
  Suppose that tree $T$ has weight $w_T$ (so $\sum_T w_T = \beta c$)
  and that we choose a tree $T$ at random with probability
  proportional to its weight.  Define the random variable $x_T$ to be
  one less than the number of edges of $T$ crossing the
  $\alpha$-minimum cut.  Note that $x_T$ is always a nonnegative
  integer.  Since we have a packing, we know that no $\alpha$-minimum
  cut edges are shared between the trees.  It follows that
\begin{align*}
\sum w_T(x_T+1) &\le \alpha c\\
\sum w_T x_T &\le \alpha c - \sum w_T\\
             &\le \alpha c - \beta c\\
\intertext{and thus}\\
E[x_T] &= \frac{1}{\sum w_T} \sum w_T x_T\\
&\le (\alpha/\beta-1)
\end{align*}
It follows from Markov's inequality that $x_T < 2$ with probability at
least $1-1/2(\alpha/\beta-1)=\frac12(3-\alpha/\beta)$.  Since $x_T$ is an
integer, $x_T<2$ implies that $x_T \le 1$, meaning that the 
 $\alpha$-minimum cut $2$-respects $T$.  

The corollary follows immediately by taking $\alpha=1$ and observing
the $\beta \ge 1/2$ by \ref{prop:trees}.
\end{proof}

\section{Combinatorics}
\label{sec:combinatorics}

As a first application of the concept of respecting constraint
trees,\marnote{later} we tighten the bounds on the number of small
cuts in a graph.  This section can be skipped without impacting the
remainder of the paper.  Our discussion is limited to {\em two-way}
cuts---that is, sets set of edges crossing a two-way vertex partition.
In contrast, the combinatorial results of~\cite{Karger:Contraction}
also apply to multiway cuts.

\begin{definition}
  A cut in $G$ is {\em $\alpha$-minimum} if its value is at most
  $\alpha$ times the minimum cut value.
\end{definition}

In~\cite{Karger:Mincut,Karger:Skeleton}, we proved that the number of
$\alpha$-minimum cuts in a graph is $O(n^{2\alpha})$.  Others have
since tightened this bound for small $\alpha$:
Benczur~\cite{Benczur:ApproxCut} bounded the number of $6/5$-minimum
cuts by $O(n^2)$, Nagamochi, Nishimura and
Ibaraki~\cite{Nagamochi:CountCuts} gave a bound of $O(n^2)$ for the
number of $4/3$-minimum cuts, and subsequently Henzinger and
Williamson~\cite{Williamson:CountCuts} showed that the number of cuts
{\em strictly less} than $3/2$ times the minimum is $O(n^2)$.  In an
unweighted $n$-vertex cycle, any even-size set of up to
$\floor{2\alpha}$ edges is the edge set of an $\alpha$-minimum cut.
Therefore, the number of $\alpha$-minimum cuts can be as large as
\begin{equation}
\binom{n}{2} + \binom{n}{4} + \cdots
+\binom{n}{\lfloor 2\alpha \rfloor} = \Theta(n^{\lfloor 2\alpha
  \rfloor}).
\label{eq:cycle}
\end{equation}
We give an upper bound that is very close to this lower bound,
matching its ``step function'' behavior 
as a function of $\alpha$.  

Our bound uses tree packings.\marnote{instead, show focus on
  unweighted graphs} It simplifies the proof to restrict attention to
unweighted tree packings---collections of edge-disjoint spanning
trees---in unweighted graphs.  The discussion of the previous section
showed that this loses no generality.  We begin with a weak but easily
proved lemma:

\begin{lemma}
  For any constant $\alpha$, there are $O(n^{\floor{2\alpha}})$
  $\alpha$-minimum cuts.
\end{lemma}
\begin{proof}
  Let $k=\floor{2\alpha}$.  Number the $\alpha$-minimum cuts from $1$
  to $\ell$.  Our goal is to bound $\ell$ by $O(n^{k})$.
  
  Recall Nash-Williams' \ref{prop:trees}~\cite{Nash:Trees} which says
  that any graph with minimum cut $c$ contains at least $c/2$
  edge-disjoint spanning trees.  Suppose that we pick one of these
  trees at random.  Let $y_i=1$ if we pick a tree that $k$-constrains
  the $i^{th}$ $\alpha$-minimum cut and $0$ otherwise.  Since this cut
  contains at most $\alpha c$ edges and each edge is in at most one of
  our $c/2$ spanning trees, the {\em expected} number of cut edges in
  our chosen spanning tree is at most $2\alpha$.  It follows from
  Markov's inequality~\cite{Motwani:RandomizedAlgorithms} and the fact
  that $1+\floor{2\alpha} > 2\alpha$ that the chosen tree has at most
  $k=\lfloor 2 \alpha \rfloor$ cut edges with constant probability.
  That is, $y_i=1$ with constant probability, meaning
  $E[y_i]=\Omega(1)$.  Thus
  \[
  E[\sum_{i=1}^{\ell} y_i] = \Omega(\sum_{i=1}^\ell 1) = \Omega(\ell).
  \]
  
  On the other hand, as discussed in the previous section, each cut
  that $k$-respects a given tree is in 1-1 correspondence with the set
  of at most $k$ edges of the tree that it cuts.  There are $O(n^k)$
  such sets of edges, so no tree can $k$-constrain more than this many
  cuts.  Thus
  \[
  E[\sum y_i] \le \max \sum y_i = O(n^k). 
  \]
  Combined with the previous bound, this means $\ell=O(n^k)$ as claimed.
\end{proof}

We now give a stronger result with a more complicated proof.

\begin{theorem}
  \label{thm:countcuts}
  The number of $\alpha$-minimum cuts is at most
  \[
  \frac{1}{\lfloor 2\alpha \rfloor + 1 -2\alpha} \binom{n}{\lfloor
      2\alpha \rfloor}(1+O(1/n)).
  \]
\end{theorem}
\begin{remark}
  For $\alpha$ a half integer, this quantity is within $1+o(1)$ of the
  lower bound in Equation~\ref{eq:cycle}.  The bound remains tight to
  within a constant factor for all values of $\alpha$ except those
  infinitesimally less than an integer (e.g., $\alpha=3-1/n$).  For
  such values, we get a better bound by considering cuts of value at
  most $\ceil{2\alpha}/2$; since this is a half-integer the theorem
  gives a bound of $\binom{n}{\ceil{2\alpha}}(1+o(1))$.
\end{remark}
\begin{proof}
  As before, we assume $G$ is unweighted and pack $c/2$ edge-disjoint
  trees in the graph.  Consider a bipartite graph $B$, with one side
  consisting of the trees in the Nash-Williams packing, and the other
  of the $\alpha$-minimum cuts.  To avoid confusion, we refer to the
  vertices of this bipartite graph as {\em nodes} and to its edges as
  {\em arcs}.  Draw an arc from a tree-node $T$ to a cut-node $C$ if
  the tree $k$-constrains the cut in $G$.  We will now assign weights
  to these arcs so that:
  \begin{enumerate}
  \item every tree node in $B$ has weighted arc degree {\em at most}
    some $\treedeg$;
  \item every cut node in $B$ has weighted arc degree {\em at least}
    some $\cutdeg$
  \end{enumerate}
  Since there are $c/2$ trees in the packing, we see from (1)
  that the total weight of arcs in $B$ is at most $c\treedeg/2$.  It 
  follows from (2) that the number of cuts in $B$ is at most
  $c\treedeg/2\cutdeg$.  We show that this quantity can be made small by an
  appropriate choice of arc weights.
  
  We now define the weight assignment.  Let $k=\floor{2\alpha}$.  For
  a given arc $(T,C)$, if tree $T$ $r$-constrains cut $C$ for some
  $r\le k$ but does not $(r+1)$-constrain $C$, then we give the arc
  weight $1+k-r$.  If more than $k$ edges of $T$ cross the cut
  $C$, we give the arc weight 0.  Since a tree can exactly $r$-constrain at
  most $\binom{n}{r}$ distinct cuts, if follows that the tree-side
  degree
  \[
  \treedeg \le \sum_{1 \le r \le k} \binom{n}{r} (1+k-r) 
  = \binom{n}{  k}(1+O(1/n)).
  \]
  Now consider the cut-side degree $\cutdeg$.  Take any cut $C$.  For
  tree $T$, let $r_T$ denote the number of edges of $T$ that cross
  $C$.  Then the weighted degree of $C$ can be written as
\begin{align*}
  \sum_{r_T \le k} (1+k-r_T) &\ge  \sum_T (1+k-r_T)\\
  &=(1+k)(c/2)-\sum_T r_T\\
  &\ge (1+k)(c/2)-\alpha c\\
  &=(1+k-2\alpha)(c/2),
\end{align*}
where the first inequality follows from the fact that all the terms
being added to the summation on the right hand side are negative, and
the second from the fact that $\sum_T r_T \le \alpha c$ since each of
the $\alpha c$ edges of the cut contribute to $r_T$ for at most one
$T$.  \comment{
\begin{equation}
  \sum_{r_T \le k} (1+k-r_T) \label{eq:treedeg}
\end{equation}
(note that this sum is only over some of the trees).  However, since
the cut $C$ has at most $\alpha c$ edges, we know that the integer values
$r_T$ satisfy the constraints
  \begin{eqnarray*}
  \sum_T r_T &\le &\alpha c\\
  r_T &\ge &0
  \end{eqnarray*}
  We now find the integer values $r_T$ that minimize the
  sum~(\ref{eq:treedeg}) subject to these constraints.  This clearly
  yields a lower bound on $\cutdeg$.
  
  Consider an assignment to $r_T$ minimizing the
  sum~(\ref{eq:treedeg}).  We can assume without loss of generality
  that $\sum_T r_T = \alpha c$ (otherwise, simply increase some $r_T$
  until the assumption is valid; this can only decrease the
  value~(\ref{eq:treedeg})).  Now suppose that for two indices $T$ and
  $T'$, we have $r_T+2 < r_{T'}$.  Then we can increment $r_T$ and
  decrement $r_{T'}$ without increasing~(\ref{eq:treedeg}) (the sum
  will remain constant unless $r_T < k+1 < r_{T'}$, in which case it
  will decrease).  It follows that without loss of generality, we can
  assume that all $r_T$ have one of two adjacent values $r$ and $r+1$.
  Since $\sum r_T = \alpha c$ and there are $c/2$ trees, we deduce
  that $rc/2\le\alpha c\le (r+1)c/2$ and thus that
  $r=\floor{2\alpha}=k$.  So let us suppose that $p$ trees are
  assigned $k$ edges while $q$ trees are assigned $k+1$ edges.  Then
  \begin{eqnarray*}
    pk + q(k+1) &= &\alpha c\\
    p+q &= &c/2
  \end{eqnarray*}
  Solving for $p$ and $q$, we find that 
  \[
  p=\frac{c}{2}(k+1-2\alpha).
  \]
  It follows that the value of~(\ref{eq:treedeg}) in this worst-case
  assignment of value $r_T$ is
  \[
  p\cdot 1+q\cdot 0=\frac{c}{2}(\floor{2\alpha}+1-2\alpha).
  \]
}
  This is a lower bound on the weight $\cutdeg$ incident on any cut-side
  node in $B$. Since we have shown above that $\treedeg \le
  \binom{n}{k}(1+O(1/n))$, it follows that our bound on the number of
  cut nodes, $c\treedeg/2\cutdeg$, is no more than the bound claimed in the
  statement of the theorem.
  
  Note that our weight assignment analysis works for any value of $k$,
  but that taking $k=\floor{2\alpha}$ gives the best ratio.  An
  argument based on linear programming duality can be used to show
  that this proof cannot be improved: no other assignment of arc
  weights can give a better bound.
\end{proof}

\subsection{Weighted Graphs}

Although our proof discussed only unweighted graphs, it clearly
extends to weighted graphs as well.  One way to see this is to redo
the proof with a weighted tree packing, but this becomes notationally
messy.  A simpler way to extend the argument to graphs with integer
edge weights is to replace a weight-$w$ edge with $w$ parallel
unweighted edges, creating an unweighted graph with the same
near-minimum cuts.  This argument can be extended to graphs with
rational edge weights as well: simply multiply all edge weights by
their least common denominator, creating an integer-weighted graph
with the same near minimum cuts.  Finally, real valued edge weights
can be handled by considering them as the limits of rational-valued
edge weight sequences.

\subsection{Discussion}

Note that a cycle has exactly $\binom{n }{ \floor{2
    \alpha}}+\cdots+\binom{n }{ 2}$ minimum cuts.  Thus for $\alpha$ a
half integer, our upper bound is tight to within $1+o(1)$.  The gap
widens when $\alpha$ is between two half-integers, reaching
$\Theta(n)$ when $\alpha$ is infinitesimally less than a half-integer.
No family of graphs has been exhibited which has more $\alpha$-minimum
cuts than the cycle, so one is tempted to conjecture that the bound
for the cycle is also the upper bound.  The only counterexample of
which we are aware is the 4-clique~\cite{Nagamochi:CountCuts}, which
has $10$ $4/3$-minimum cuts (each singleton or pair of vertices) as
compared to the $6$ of a 4-cycle.  But perhaps this is a unique exception.

Another interesting open question regards the number of cuts {\em
  strictly less} than $\alpha$ times the minimum.  For $\alpha$ a half
integer, our above theorem only gives a bound of
$O({n}^{\ceil{2\alpha}})$; a bound of
$O({n}^{\floor{2\alpha}})$, as exhibited by the cycle, seems more
plausible.  Such a bound has been proven for
$\alpha=3/2$~\cite{Williamson:CountCuts}.

\section{Algorithms for Finding Good Trees}
\label{sec:packing}

We now apply the ideas of the previous combinatorial sections
algorithmically.  We prove that in any graph, we can find quickly a
small set of trees such that the minimum cut $2$-respects some of
them.  Therefore, we can find the minimum cut by enumerating only the
cuts that $2$-respect these few trees.  

\subsection{Packing Algorithms}

Finding the trees is in a sense trivial from the preceding
discussion, which showed that a maximum tree packing had the right
property.  Unfortunately, maximum tree packings in undirected graphs
are hard to find.  Gabow and Westermann~\cite{Gabow:Packing} gave an
algorithm for unweighted graphs that runs in
$\Olog(\min(mn,m^2/\sqrt{n}))$ time.
Barahona~\cite{Barahona:TreePacking} gives an algorithm with an
$\Olog(mn)$ running time for weighted graphs. Both of
these running times are dramatically larger than the one we want to
achieve, so we will use two other approaches that find non-maximum tree
packings.

One approach is due to Gabow.  Although it does not yield a maximum
packing, it yields a packing of value $c/2$---sufficient for out
purposes.  Gabow's algorithm is actually an algorithm for directed
graphs: in such a graph, it finds the $c$ directed spanning trees
guaranteed by Edmonds' theorem~\cite{Edmonds:Branchings} in
$O(mc\log(n^2/m))$ time.  We can use this algorithm by turning each of
our undirected graph's edges into two edges, one in each direction.
This gives us a directed graph with minimum cut $c$.  Gabow's
algorithm finds a packing of $c$ directed-edge disjoint trees in this
graph.  If we now ignore edge directions, we get a set of $c$ trees
such that each of our undirected edges is in at most $2$ trees.  It
follows that if we give each of our trees a weight of $1/2$, we have a
packing of value $c/2$ to which our previous arguments
(\ref{prop:packing}) apply.

An alternative approach is due to Plotkin, Shmoys, and
Tardos~\cite{Plotkin:Packing}.  They give an algorithm that packs
spanning trees by repeatedly adding one weighted tree to the packing.
Which tree to add and the weight it is given are determined by a
minimum-cost spanning tree computation, using costs determined by the
current packing.  After $\Olog(c/\epsilon^2)$ iterations (which take
time $\Olog(mc/\epsilon^2)$), the packing has value at least
$(1-\epsilon)$ times the maximum, and thus at least
$(1-\epsilon)c/2$.  Although it is more
complicated that Gabow's algorithm in that it works with fractional
value, this method has the
 attraction of working only with undirected spanning trees,
emphasizing that our algorithm is able to avoid any reliance on
directed-graph algorithms.  It will also be useful when we develop
parallel algorithms.

\subsection{Sampling}

Unfortunately, both of the above algorithms have running times
dependent on $c$, so using them directly is unacceptable.  However,
we now show use a random sampling step to reduce the effective minimum
cut in the graph to $O(\log n)$, which means that both schemes can run
in $\Olog(m)$ time.

\begin{theorem}\labelprop{Theorem}{prop:some trees}
  Given any weighted undirected graph $G$, in $O(m+n\log^3 n)$ time we
  can construct a set of $O(\log n)$ spanning trees such that the
  minimum cut $2$-respects $1/3$ of them with high probability.
\end{theorem}
\begin{proof}
  In a previous paper~\cite{Karger:Skeleton}, we showed how to
  construct, in linear time for any $\epsilon$, a {\em
    skeleton} graph $H$ on the same vertices with the following
  properties:
  \begin{itemize}
  \item $H$ has $m' = O(n\epsilon^{-2}\log n)$ edges,
  \item the minimum cut of $H$ is $c' = O(\epsilon^{-2}\log n)$,
  \item the minimum cut in $G$ corresponds (under the same vertex
    partition) to a $(1+\epsilon)$-times minimum cut of $H$.
  \end{itemize}

  We already know that any tree packing of value $c'/2$ in the
  skeleton $H$ will have many trees that 2-constrain the minimum cut
  in $H$.  Intuitively, for small $\epsilon$, since the minimum cut in
  $G$ is a near-minimum cut in $H$, the tree packing will also contain
  many trees that $2$-constrain this near minimum cut.  We now
  formalize this intuition.
  
  Set $\epsilon=1/6$ in the skeleton construction.  Since $H$ has
  minimum cut $c'$, Gabow's algorithm can be used to find a packing in
  $H$ of weight $c'/2$ in $O(m'c'\log n)=O(n\log^3 n)$ time.  The
  original minimum cut of $G$ has at most $(1 +\epsilon)c'$ edges in
  $H$, so by \ref{prop:packing}, a fraction $\frac12(1-2\epsilon)=1/3$
  of the trees $2$-constrain this cut in $H$.  But this
  $(1+\epsilon)$-minimum cut of $H$ has the same vertex partition as
  the minimum cut of $G$, implying that the same trees $2$-constrain
  the minimum cut of $G$.
\end{proof}

\begin{remark}
  The randomized construction of the skeleton in this proof is the
  only step of our minimum cut algorithm that involves randomization.
  A deterministic replacement of \ref{prop:some trees} would yield a
  deterministic minimum cut algorithm.
\end{remark}

An alternative construction based on the Plotkin-Shmoys-Tardos
algorithm can also be applied to the sample.  We can use that
algorithm to find a tree packing of value $(1-\delta)c'/2$ in
$O(n\log^3 n)$ time for any constant $\delta$.  \ref{prop:packing}
again shows that for $\delta = \epsilon= 1/6$, say, a $1/10^{th}$
fraction of the packed weight $2$-constrains the
$(1+\epsilon)$-minimum cut in $H$ that corresponds to the minimum cut
in $G$.  Thus, if we choose a random tree (with probability
proportional to its weight), there is a constant probability that it
$2$-constrains the minimum cut.  Performing $O(\log n)$ such random
selections picks at least one tree that $2$-constrains the minimum cut
with high probability (In fact, since the algorithm only packs
$O((\log n)/\delta^2)$ distinct trees we can even afford to try them
all).  This gives another $O(m+n\log^3 n)$-time algorithm for finding
a good set of trees.  It also has the following corollary that will be
applied to our parallel minimum cut algorithm.

\begin{corollary}
  \labelprop{Corollary}{prop:some trees parallel} In $\RNC$ using
  $m+n\log n$ processors and $O(\log^3 n)$ time, we can find a set of
  $O(\log n)$ spanning trees such that with high probability, a
  constant fraction of them (by weight) $2$-constrain the minimum cut.
\end{corollary}
\begin{proof}
  The skeleton construction can be performed as before (and is trivial
  to parallelize~\cite{Karger:Skeleton}).  We have just argued that
  the algorithm of~\cite{Plotkin:Packing} can find an approximately
  maximum packing in the skeleton using $O(\log^2 n)$ minimum spanning
  tree computations.  Minimum spanning trees can be found in parallel
  using $m'$ processors and $O(\log n)$ time~\cite{Johnson:MST}, and
  the other operation of~\cite{Plotkin:Packing} are trivial to
  parallelize with the same time and processor bounds.  Thus the
  claimed bounds follow.
\end{proof}

The remainder of this paper is devoted to the following question:
given a tree, find the minimum cut that $2$-respects it.  Applying the
solution to the $O(\log n)$ trees described by the previous lemma
shows that we can find the minimum cut with high probability.  This gives
the following lemma:

\begin{lemma}
\labelprop{Lemma}{prop:paradigm}
Suppose the minimum cut that 2-respects a given tree can be found in
$T(m,n)$ time.  Then the minimum cut of a graph can be found with
constant probability in $T(m,n)+O(m+n\log^3 n)$ time and with high
probability in $O(T(m,n)\log n+m+n\log^3 n)$ time.
\end{lemma}
\begin{proof}
  We have seen above that a constant fraction (by weight) of the trees
  in our skeleton packing 2-constrain the minimum cut.  So choosing a
  tree at random and analyzing it yields the minimum cut with constant
  probability.  Trying all $O(\log n)$ trees identifies the minimum
  cut with high probability so long as the skeleton construction
  worked, which happens with high probability.
\end{proof}

We give two algorithms for analyzing a tree.  The first algorithm (in
Section~\ref{sec:tree 2 slow}) uses the fact that a tree contains only
$\binom{n}{2}$ pairs of edges, one of which defines the minimum cut
$2$-respecting the tree.  A very simple dynamic programming step is
used to compute the cut values defined by all $\binom{n}{ 2}$ pairs of
edges in $O(n^2)$ time.  The second algorithm (in
Section~\ref{sec:tree 2 fast}) aims for a linear-time bound, and must
therefore avoid enumerating all pairs of edges.  We describe local
optimality conditions showing that only certain pairs of edges can
possibly be the pair defining the minimum cut, and give an algorithm
that runs quickly by only trying these plausible pairs.

\section{Minimum Cuts that $1$-Respect a Tree}
\label{sec:tree 1 slow}

To introduce our approach to analyzing a particular tree, we consider
a simple special case: a tree that {\em 1-constrains} the minimum cut.
In such a tree there is one tree edge such that, if we remove it, the
two resulting connected subtrees correspond to the two sides of the
minimum cut.  This section is devoted to a proof of the following:

\begin{lemma}
  \label{lem:tree 1 slow}
  The values of all cuts that $1$-respect a given spanning tree can be
  determined in $O(m+n)$ time.
\end{lemma}

\begin{corollary}
  The minimum cut that $1$-respects a given spanning tree can be found
  in $O(m+n)$ time.
\end{corollary}

We have already seen that being able to identify all cuts that
2-respect a tree will let us find the minimum cut.  Later, we will
exhibit a class of graphs for which finding the minimum 1-respecting
cut will suffice for finding the minimum cut.

We begin with some definitions.  Suppose that we root the tree at an
arbitrary vertex.

\begin{definition}
  $\descendants{v}$ is the set of vertices that are descendants
  of $v$ in the rooted tree, including $v$.
\end{definition}

\begin{definition}
  $\ancestors{v}$ is the set of vertices that are ancestors of
  $v$ in the rooted tree, including $v$.
\end{definition}
Note that $\descendants{v}\cap\ancestors{v}=v$.

\begin{definition}
  $\crossweight{X}{Y}$ is the total weight of edges crossing from
vertex set $X$ to vertex set  $Y$.
\end{definition}

In particular, $\crossweight{v}{w}$ is the weight of the edge $(v,w)$ if it
exists, and $0$ otherwise.

\begin{definition}
  $\cutval{S}$ is the value of the cut whose one side is vertex set
  $S$, \ie\ $\crossweight{S}{\overline{S}}.$
\end{definition}

Once we have rooted the tree, the cuts that $1$-respect
the tree have the form $\cutval{\descendants{v}}$ for vertices $v$.
Using this observation, we now prove Lemma~\ref{lem:tree 1 slow}.

\subsection{$1$-respecting a path}
\label{sec:path-1}

As a first step, suppose the tree is in fact a path $v_1,\ldots,v_n$
rooted at $v_1$.  We compute all values $\cutval{\descendants{v_i}}$
in linear time using a dynamic program.  First compute
$\crossweight{v_i}{\ancestors{v}_i}$ and
$\crossweight{v_i}{\descendants{v}_i}$ for each $v_i$; this takes one
$O(m)$-time traversal of the vertices' adjacency lists.  We now claim 
that the following recurrence applies to the cut values:
\begin{eqnarray*}
  \cutval{\descendants{v_n}} &= &\crossweight{v_n}{\ancestors{v}_n}\\ 
  \cutval{\descendants{v_i}} &=
  &\cutval{\descendants{v}_{i+1}}+\crossweight{v_i}{\ancestors{v}_i}
  -\crossweight{v_i}{\descendants{v}_i}
\end{eqnarray*}
This recurrence follows from the fact that when we move $v_i$ from
below the cut to above it, the edges from $v_i$ to its descendants
become cut edges while those from $v_i$ to its ancestors stop being
cut edges.  It follows that all $n$ cut values can be computed in
$O(n)$ time by working up from $v_n$.

\subsection{$1$-respecting a tree}

We now extend our dynamic program from paths to general trees.

\begin{definition}
  Given a function $f$ on the vertices of a tree, the {\em treefix sum}
  of $f$, denoted $\treefix{f}$, is the function
  \[
  \treefix{f}(v) = \sum_{w \in\descendants{v}} f(w).
  \]
\end{definition}

\begin{lemma}
  \label{lem:treefix}
  Given the values of a function $f$ at the tree nodes, all values of
  $\treefix{f}$ can be computed in $O(n)$ time.
\end{lemma}
\begin{proof}
  Perform a postorder traversal of the nodes.  When we visit a node
  $v$, we already will have computed (by induction) the values at each
  of its children. Adding these values takes time proportional to the
  degree of $v$; adding in $f(v)$ gives us $\treefix{f}(v)$.  Thus,
  the overall computation time is proportional to the sum of the node
  degrees, which is just the number of tree edges, namely $n-1$.
\end{proof}

We now compute the values $\cutval{\descendants{v}}$ via treefix sums.
Let $\degree{v}$ denote the (weighted) degree of vertex $v$.  Let
$\lcaweight{v}$ denote the total weight of edges whose endpoints'
least common ancestor is $v$.

\begin{lemma}
  $
  \cutval{\descendants{v}} = \degreesum{v}-2\bothbelow{v}.
  $
\end{lemma}
\begin{proof}
  The term $\degreesum{v}$ counts all the edges leaving descendants of
  $v$.  This correctly counts each edge crossing the cut defined by
  $\descendants{v}$, but also double-counts all edges with both
  endpoints descended from $v$.  But an edge has both endpoints
  descended from $v$ if and only if its least common ancestor is in
  $\descendants{v}$.  Thus the total weight of such edges is
  $\bothbelow{v}$.
\end{proof}

Now note that the functions $\degree{v}$ and $\lcaweight{v}$ can both
be computed for all $v$ in $O(m)$ time. Computing $\degreel$ is trivial.
Computing $\bothbelowl$ is trivial if we know the least common
ancestor of each edge, but these can be determined in $O(m)$
time~\cite{Gabow:LCA,Berkman:LCA,Schieber:LCA}.  From these
quantities, according to Lemma~\ref{lem:treefix}, two $O(n)$-time
treefix computations suffice to determine the minimum cut.  This
proves Lemma~\ref{lem:tree 1 slow}.

In Section~\ref{sec:fat}, we will describe a class of graphs for which
this ``1-respects'' test is sufficient to find the minimum cut.

\section{Minimum Cuts that $2$-respect a Tree in $O(n^2)$ Time}
\label{sec:tree 2 slow}

We now extend our dynamic program to find the minimum cut that
$2$-respects a given tree in $O(n^2)$ time.  This will immediately
yield an $O(n^2\log n)$-time algorithm for finding all minimum cuts.
The dynamic program can be parallelized, yielding an $n^2$-processor
$\RNC$ algorithm for the problem.  The trace of the dynamic program
also provides a small-space data structure representing all the
near-minimum cuts.

\subsection{A sequential algorithm}

This section is devoted to proving the following:

\begin{lemma}
  \label{lem:tree 2 slow}
  The values of all cuts that $2$-respect a given tree can be found in
  $O(n^2)$ time.
\end{lemma}

A sequential algorithm is an immediate corollary (using
\ref{prop:paradigm}):

\begin{theorem}
  \label{thm:mincut slow}
  There is a Monte Carlo algorithm that finds a minimum cut with
  constant probability in $O(n^2)$ time and finds all minimum cuts
  with high probability in $O(n^2\log n)$ time.
\end{theorem}

We now prove the lemma.  After rooting the given tree and applying the
algorithm of Section~\ref{sec:tree 1 slow} in case some tree
1-constrains the minimum cut, we can assume that (for the right tree)
exactly 2 edges cross the minimum cut.  These two edges are the parent
edges of two vertices $v$ and $w$.

\begin{definition}
  We say vertices $v$ and $w$ are {\em incomparable}, writing
  $\incomparable{v}{w}$, if $v \notin \descendants{w}$ and $w \notin
  \descendants{v}$.  In other words, $v$ and $w$ incomparable if they
  are not on the same root-leaf path.
\end{definition}

Suppose the minimum cut that $2$-respects the tree cuts the parent
edges of vertices $v$ and $w$.  If $\incomparable{v}{w}$, then the
parent edges of $v$ and $w$ define a cut with value
$\cutval{\descendants{v} \cup \descendants{w}}$.  If (without loss of
generality) $v \in \descendants{w}$, then their parent edges define a
cut with value $\cutval{\descendants{w}-\descendants{v}}$.  We start
with the first case.  Assuming $\incomparable{v}{w}$,
\[
\cutval{\descendants{v}\cup\descendants{w}} =
\cutval{\descendants{v}}+\cutval{\descendants{w}} -
2\crossweight{\descendants{v}}{\descendants{w}}
\]
because $\cutval{\descendants{v}}$ and $\cutval{\descendants{w}}$ each
include the edges of $\crossweight{\descendants{v}}{\descendants{w}}$
that should not be counted in
$\cutval{\descendants{v} \cup \descendants{w}}$.  Since the values
$\cutval{\descendants{v}}$ are already computed for all $v$, we need
only compute $\crossweight{\descendants{v}}{\descendants{w}}$ for
every $v$ and $w$.  This can be done using treefix sums as in the
previous section.  First, let
\[
f_v(w) = \crossweight{v}{w}
\]
be the weight of the edge connecting $v$ to $w$, or $0$ if there is
no such edge.  It follows that
\[
\treefix{f}_v(w) = \crossweight{v}{\descendants{w}}.
\]
Therefore, we can determine all $n^2$ values
\[
g_w(v) = \crossweight{v}{\descendants{w}} = \treefix{f}_v(w)
\]
in $O(n^2)$ time by performing $n$ treefix computations, one for each
$f_v$.  At this point, since
\[
\treefix{g_w}(v) = \crossweight{\descendants{v}}{\descendants{w}},
\]
we can determine all the desired quantities with another $n$ treefix
sums, one for each $g_w$.

Now consider the second case of $v \in \descendants{w}$.  In this
case, 
\begin{eqnarray*}
  \cutval{\descendants{w}-\descendants{v}}
  &=&\cutval{\descendants{w}}-\cutval{\descendants{v}}
  +2\crossweight{\descendants{v}}{\descendants{w}-\descendants{v}},
\end{eqnarray*}
for subtracting $\cutval{\descendants{v}}$ from
$\cutval{\descendants{w}}$ correctly subtracts the edges connecting
$\descendants{v}$ to $\ancestors{w}$, but also incorrectly subtracts
all the edges running from $\descendants{v}$ to
$\descendants{w}-\descendants{v}$.  But we claim that
\[
\crossweight{\descendants{v}}{\descendants{w}-\descendants{v}}
=
\treefix{g_w}(v)-2\bothbelow{v}.
\]
For $\treefix{g_w}(v)$ counts every edge with one endpoint in
$\descendants{v}$ and the other in $\descendants{w}$, which includes
everything we want to count but also incorrectly counts (twice) all
edges with both endpoints in $\descendants{v}$.  This is cancelled by
the subtraction of $2\bothbelow{v}$.

Thus, once we have computed $g_w$, computing the cut value for each
pair $v \in \descendants{w}$ in $O(n^2)$ time is trivial.

\subsection{A parallel algorithm}
\label{sec:parallel}

We can parallelize the above algorithms.  Note that they involve two
main steps: finding a proper packing of trees and finding the minimum
cut constrained by each tree.  Recall from \ref{prop:some trees
  parallel} that the algorithm of~\cite{Plotkin:Packing} can be used
to find a satisfactory packing in $O(\log^3 n)$ time using $m+n\log n$
processors.  So we need only parallelize the above algorithms for a
particular tree.  But the only computations performed there are least
common ancestor lookups, which can be performed optimally in parallel
by, e.g., the algorithm of Schieber and Vishkin~\cite{Schieber:LCA};
and treefix sum computations, which can also be performed optimally in
parallel~\cite[Section 2.2.3]{Karp:Parallel}.

\begin{corollary}
  Minimum cuts can be found in $\RNC$ with high probability in
  $O(\log^3 n)$ time using $n^2/\log^2 n$ processors for general
  graphs.\marnote{ and using $m$ processors for $\epsilon$-fat graphs.}
\end{corollary}

\subsection{Data Structures}

Previously, a linear-space representation of all minimum cuts was
known~\cite{Dinitz:Cactus}, but the best representation of
approximately minimum cuts for general $\alpha$ required $\Theta(n^{2\alpha})$
space~\cite{Karger:Contraction}.  Benczur~\cite{Benczur:ApproxCut}
gave an $O(n^2)$-space representation for $\alpha < 6/5$.    We can now do
better.

\begin{theorem}
  Given a graph with $k$ $\alpha$-minimum cuts, $\alpha < 3/2$, there
  is a data structure that represents them in $O(k+n\log n)= O(n^2)$
  space that can be constructed in $O(n^2\log n)$ time or in $\RNC$
  using $n^2$ processors.  The value of  a near-minimum cut can be
  looked up in $O(n)$ time.
\end{theorem}
\begin{proof}
  The output of our $\Olog(n^2)$-time algorithm can serve as a data
  structure for representing the near-minimum cuts that $2$-respect
  the tree it analyzes.  We simply list the pairs of edges that
  together induce a near minimum cut.  Now consider any $\alpha <
  3/2$.  We know that with high probability every $\alpha$-minimum cut
  $2$-respects one of the $O(\log n)$ trees we inspect and will
  therefore be found.

  Notice that given a cut (specified by a vertex partition) we can use
  this data structure to decide if it is $\alpha$-minimum, and if so
  determine its value, in $O(n\log n)$ time: simply check, for each of
  the $O(\log n)$ trees in the data structure, whether at most 2 tree
  edges cross the cut.  If the cut is a small cut, there will be some
  tree for which this is true.  We will have recorded the value of the
  cut with the edge pair (or singleton) that crosses it.  

We can improve this query time to $O(n)$ if we use a perfect hash
  function to map each vertex set that defines a small cut to a
size-$O(k)$ table that says which tree that particular small cut
$2$-respects.  Given a query (vertex set) we can map that vertex set
into the hash table and check the one tree indicated by the
appropriate table entry.  
\end{proof}

In the sequential case, this data structure can actually be
constructed deterministically if we are willing to take some extra
time.  Simply perform the initial tree packing in the original graph
rather than the skeleton.  By \ref{prop:packing}, we know that any one
minimum cut $2$-respects at least half the trees of this packing.
Thus, if we pick a random tree, we expect it to $2$-constrain half the
minimum cuts.  It follows that some tree $2$-constrains half the
minimum cuts. We can find such a tree {\em deterministically} via an
exhaustive search.  First, enumerate all the small cuts using, for example, the
deterministic algorithm of~\cite{Vazirani:EnumerateCuts}.  Then, check
each tree to find one that 2-constraints half the minimum cuts.  In
the case of an unweighted graph, we must check $c<m$ trees.  In a
weighted graph, we can start from a maximum packing with a polynomial
number of distinct trees in it~\cite{Barahona:TreePacking} so that we
can check them all in polynomial time.  We now recursively find trees
that $2$-constrain the remaining minimum cuts that the first tree missed.
Since we halve the number of remaining min-cuts each time, we only
need $\log_2 n^{2\alpha}$ trees.

We can easily extend this approach to represent $\alpha$-minimum cuts
for any constant $\alpha$.  Simply observe that \ref{prop:some trees}
generalizes to argue that of the $O(\log n)$ trees we find there, a
constant fraction will $\floor{2\alpha}$-respect any given
$\alpha$-minimum cut with high probability.  We can therefore
represent all $\alpha$-minimum cuts by considering all sets of
$\floor{2\alpha}$ edges in each tree.  For any one tree, the values
induced by all sets of $\floor{2\alpha}$ edges can be found in
$O(n^{\floor{2\alpha}})$ time using treefix computations as above.

\begin{lemma} 
  There is a data structure that represents all $k$ of the
  $\alpha$-minimum cuts in a graph using $O(n\log n+k\alpha)=O(\alpha
  n^{\floor{2\alpha}})$ space.  Its randomized construction requires
  $O(n^{\floor{2\alpha}})$ time and space or $O(n^{\floor{2\alpha}})$
  processors in $\RNC$.  It can also be constructed deterministically
  in polynomial time.
\end{lemma}

\section{Near-Linear Time}
\label{sec:tree 2 fast}

We now reconsider the algorithm of Section~\ref{sec:tree 2 slow} and
show how its objective can be achieved in $O(m\log^2 n)$ time on any
given tree.   In this section we prove the following lemma.

\begin{lemma}
  \labelprop{Lemma}{prop:tree 2 fast} The minimum cut that 2-respects a tree
  can be found in $O(m\log^2 n)$ time.
\end{lemma}

Applying \ref{prop:paradigm} yields our main theorem:

\begin{theorem}
  \label{thm:mincut fast}
  The minimum cut of a graph can be found with constant probability in
  $O(m\log^2 n+n\log^3 n)$ time or with high probability in $O(m\log^3
  n)$ time.
\end{theorem}

To prove the lemma, as in Section~\ref{sec:tree 2 slow}, we first
consider finding the pair minimizing $\cutvals{v}{w}$ for
$\incomparable{v}{w}$; the almost identical case of minimizing
$\cutval{\descendants{w}-\descendants{v}}$ for $v \in \descendants{w}$
is deferred to the end of the section.  Rather than explicitly
computing $\cutvals{v}{w}$ for all pairs $v$ and $w$, which clearly
takes $\Theta(n^2)$ time, we settle for finding for each $v$ a vertex
$w$ that minimizes $\cutvals{v}{w}$.  This will clearly let us
identify the minimum cut.

\subsection{Precuts}

Recall that for any vertices $\incomparable{w}{v}$,
\[
\cutvals{v}{w} = \cutval{\descendants{v}}+\cutval{\descendants{w}}
-2\crossweight{\descendants{v}}{\descendants{w}}.
\] 
We now factor out the contribution of $\cutval{\descendants{v}}$ to
this quantity: 

\begin{definition}
  \label{def:precut}
  The {\em $v$-precut at $w$,} denoted $\precut{v}{w}$, is the value
  \[
  \precut{v}{w}= \cutvals{v}{w}-\cutval{\descendants{v}} =
  \cutval{\descendants{w}} - 2
  \crossweight{\descendants{v}}{\descendants{w}}
  \]
  if $\incomparable{w}{v}$ and $\infty$ otherwise.
\end{definition}

\begin{definition}
  \label{def:minprecut}
  The {\em minimum $v$-precut}, denoted $\minprecut{v}$, is the value
  \[
  \min \set{ \precut{v}{w}
    \mid \exists (v',w')\in E,\ 
    v' \in \descendants{v}, \ w' \in \descendants{w}}
  \]
\end{definition}

The point of this somewhat odd definition is to notice that for a
given $v$, only certain vertices $w$ are candidates for forming the
minimum cut with $v$, as is seen in the following lemma.

\begin{lemma}
  \label{lem:precut}
  If it is defined by incomparable vertices, the minimum cut is
  $\min_v (\cutval{\descendants{v}}+\minprecut{v})$.
\end{lemma}
\begin{proof}
  Let the minimum cut be induced by $v$ and a vertex
  $\incomparable{w}{v}$.  It suffices to show that $w$ is in the set
  over which we minimize to define $\minprecut{v}$ in
  Definition~\ref{def:minprecut}.  But this follows from the fact that
  each side of the minimum cut induces a connected subgraph of $G$,
  for this implies that there must be an edge $(v',w')$ from
  $\descendants{v}$ to $\descendants{w}$.  To see that each side of
  the minimum cut must be connected, note that if one side of the
  minimum cut has two disconnected components then we can reduce the
  cut value by moving one of these components to the other side of the
  cut.
\end{proof}

Suppose that we have already used the linear-time procedure of
Section~\ref{sec:tree 1 slow} to compute the values
$\cutval{\descendants{v}}$ for every $v$.  It follows from
Lemma~\ref{lem:precut} that it suffices to compute $\minprecut{v}$ for
every vertex $v$: the minimum cut can then be found in $O(n)$ time.
We now show how these minimum precuts can be computed in $O(m\log^2 n)$
time on any tree.  We first show that we can find $\minprecut{v}$ for
a leaf $v$ in time proportional to its degree.  We then extend this
approach to handle a {\em bough}---a path rising from a leaf until it
reaches a vertex with more than one child.  Finally, we handle general
trees by repeatedly pruning boughs.

We begin by giving each vertex $w$ a variable $\marker{w}$.  While
working on vertex $v$, this
variable will be used to accumulate $\precut{v}{w}$.  Initially, we set
$\marker{w} = \cutval{\descendants{w}}$.  

\subsection{A Leaf}

First consider a particular leaf $v$.  To compute
$\minprecut{v}$, consider the following procedure:
\begin{enumerate}
\item \label{item:subtract} For each vertex $w$, subtract
  $2\crossweight{v}{\descendants{w}}$ from $\marker{w}$ (so that
  $\marker{w} = \precut{v}{w}$).
\item \label{item:minimize} $\minprecut{v}$ is the minimum of
  $\marker{w}$ over all $\incomparable{w}{v}$ that are ancestors of a
  neighbor of $v$.
\end{enumerate}
The correctness of this procedure follows from Lemma~\ref{lem:precut},
the initialization of $\marker{w}$, and the definition of
$\minprecut{v}$.

To implement this procedure efficiently, we use the {\em dynamic tree}
data structure of Sleator and Tarjan~\cite{Sleator:DynamicTrees}.
Given a tree, this structure supports (among others) the following
operations on a node $v$:
\begin{description}
\setlength{\itemindent}{\parindent}
\item[$\addpath{v}{x}$:] add $x$ to $\marker{u}$
  for every $u \in \ancestors{v}$.
\item[$\minpath{v}$:] return 
  $\displaystyle
  \min_{u \in \ancestors{v}} \marker{u}
  $
  as well as the $u$ achieving this minimum.
\end{description}
Each such dynamic tree operation takes $O(\log n)$ amortized time
steps.

We use dynamic trees to compute $\minprecut{v}$.  Recall that
$\crossweight{v}{u}$ is the weight of the edge connecting vertices $v$
and $u$, or $0$ if no such edge exists.  Apply procedure ${\proc
  LocalUpdate}$ from Figure~\ref{fig:localupdate}.

\begin{figure}[htbp]
  \boxtext{
    \procti{LocalUpdate$(v)$}
    \bigskip
    \begin{enumerate}
    \item \label{step:ancestors} Call $\addpath{v}{\infty}$.
    \item \label{step:add} For each edge $(v,u)$, 
      call $\addpath{u}{-2\edgeweight{v}{u}}$.
    \item \label{step:min} For each edge $(v,u)$, call $\minpath{u}$
    \item Return the minimum result of Step~\ref{step:min}.
    \end{enumerate}}
  \label{fig:localupdate}
  \caption{procedure ${\proc LocalUpdate}$}
\end{figure}

\begin{lemma}
  \label{lem:leaf}
After Step~\ref{step:add} in a call to ${\proc LocalUpdate}(v)$, every
  $w$ has $\marker{w}=\precut{v}{w}$.
\end{lemma}
\begin{proof}
  Step~\ref{step:ancestors} assigns an infinite $\marker{w}$ to
  every ancestor $w$ of $v$, as required in the definition of
  $\precut{v}{w}$.  Next, $2\edgeweight{v}{u}$ is subtracted in
  Step~\ref{step:add} from every $\marker{w}$ such that $u \in
  \descendants{w}$.  Therefore, the total amount subtracted from
  $\marker{w}$ is $2\crossweight{v}{\descendants{w}}$, as required in
  Definition~\ref{def:precut}.  It follows that after
  Step~\ref{step:add}, $\marker{w} = \precut{v}{w}$.
\end{proof}

\begin{lemma}
  \label{lem:leaf2}
  ${\proc LocalUpdate}(v)$ called on a leaf $v$ returns $\minprecut{v}$.
\end{lemma}
\begin{proof}
  Step~\ref{step:min} minimizes over $\incomparable{w}{v}$ that are
  ancestors of neighbors of $v$.  According to
  Definition~\ref{def:minprecut} this will identify $\minprecut{v}$.
\end{proof}

It follows that for a leaf $v$ with $d$ incident edges, we can find
$\minprecut{v}$ via $O(d)$ dynamic tree operations that require
$O(d\log n)$ time.

\subsection{A Bough}

We generalize the above approach.  Let a {\em bough} be a
maximal path starting at a leaf and traveling upwards until it reaches
a vertex with more than one child (this vertex is not in the bough).

\begin{lemma} \label{lem: update precut}
  Let $v$ be a vertex with a unique child $u$.  Then either
  $\minprecut{v}=\minprecut{u}$, or else $\minprecut{v}=\precut{v}{w}$
  for some ancestor $w$ of a neighbor of vertex $v$.
\end{lemma}
In other words, given the value $\minprecut{u}$, we need only ``check''
ancestors of neighbors of $v$ to determine $\minprecut{v}$.
\begin{proof}
  We know that $\minprecut{v}=\precut{v}{w}$ for a certain $w$.  Suppose
  $w$ is not an ancestor of a neighbor of $v$.  Then there is no edge
  from $v$ to $\descendants{w}$, meaning
  $\crossweight{\descendants{v}}{\descendants{w}} =
  \crossweight{\descendants{u}}{\descendants{w}}$.  It follows that
  $\minprecut{u} \le \precut{u}{w} =\precut{v}{w}=\minprecut{v}$.  But
  $u \in \descendants{v}$ implies that for any $x$,
  $\crossweight{\descendants{u}}{\descendants{x}} \le
  \crossweight{\descendants{v}}{\descendants{x}}$, so we know that
  $\minprecut{u} \ge \minprecut{v}$.  Therefore
  $\minprecut{u}=\minprecut{v}$.
\end{proof}

We use the above lemma in an algorithm for processing an entire bough.
Given a bough with $d$ edges incident in total on its vertices, we
show how to process all vertices in the bough in $O(d\log n)$ time.
We use the recursive procedure ${\proc MinPrecut}$ from
Figure~\ref{fig:minprecut}.  This procedure is initially called on the
topmost vertex in the bough.  Although we have formulated the
procedure recursively to ease the proof exposition, it may be more
natural to unroll the recursion and think of the algorithm as
executing up from a leaf.

\begin{figure}[htbp]
  \boxtext{
    \procti{MinPrecut$(v)$}\\
    \begin{tabbing}
      else \= else \= else \= \kill
      {\bf if} $v$ is the leaf of the bough\\
      \> call ${\proc LocalUpdate}(v)$ (also computes $\minprecut{v}$)\\
      {\bf else}\\
      \> Let $u$ be the child of $v$\\
      \> $c_1 \leftarrow {\proc MinPrecut}(u)$ (updates some
$\marker{\cdot}$ entries)\\
      \> $c_2 \leftarrow {\proc LocalUpdate}(v)$ \\
      \> {\bf return} $\min (c_1,c_2)$
    \end{tabbing}
    }
  \label{fig:minprecut}
  \caption{procedure ${\proc MinPrecut}$}
\end{figure}

We claim that ${\proc MinPrecut}$ computes the desired quantity.  To
prove this, we prove a stronger statement by induction:

\begin{lemma}
  A call to ${\proc MinPrecut}(v)$
  \begin{enumerate}
  \item \label{item:values} sets $\marker{w} = \precut{v}{w}$ for each
    $w$, and
  \item \label{item:returns} returns $\minprecut{v}$.
  \end{enumerate}
\end{lemma}
\begin{proof}
  By induction.  We first prove Claim~\ref{item:values}.  The base case
  is for $v$ a leaf, and follows from our analysis of {\proc
    LocalUpdate} for a leaf in Lemmas~\ref{lem:leaf}
  and~\ref{lem:leaf2}.  Now suppose $v$ has child $u$.  After calling
  ${\proc MinPrecut}(u)$, we know by induction that
  $\marker{w}=\precut{u}{w}$. Thus, after we
  execute ${\proc LocalUpdate}(v)$, which (as shown in Lemma~\ref{lem:leaf})
  decreases each $\marker{w}$ by $\crossweight{v}{\descendants{w}}$, entry
 $\marker{w}$ is updated to be
  \begin{eqnarray*}
\precut{u}{w} -
    2\crossweight{v}{\descendants{w}} &=
    &(\cutval{\descendants{w}}-2\crossweight{\descendants{u}}{\descendants{w}})
    -2\crossweight{v}{\descendants{w}}\\ &=
    &\cutval{\descendants{w}}-2\crossweight{v \cup
      \descendants{u}}{\descendants{w}}\\ &=
    &\cutval{\descendants{w}}-2\crossweight{\descendants{v}}{\descendants{w}}\\
    &= &\precut{v}{w} 
  \end{eqnarray*}

  We now prove Claim~\ref{item:returns}.  Suppose $v$ has child $u$.
  According to Lemma~\ref{lem: update precut}, there are two
  possibilities for $\minprecut{v}$.  One possibility is that
  $\minprecut{v}=\minprecut{u}$.  But by induction, the recursive call
  to ${\proc MinPrecut}(u)$ sets $c_1=\minprecut{u}$.  On the other
  hand, if $\minprecut{v}\ne \minprecut{u}$, then by Lemma~\ref{lem:
    update precut}, $\minprecut{v}=\precut{v}{w}$ for an ancestor of a
  neighbor of $v$.  But then $c_2$ is set to $\minprecut{v}$ when we
  call ${\proc LocalUpdate}(v)$.  In either case, the correct value is
  $\min (c_1,c_2)$.
\end{proof}

Based on this lemma, the correctness of the algorithm for finding all
minimum precut values in a bough is immediate.

\subsection{A Tree}

We now use the bough algorithm repeatedly to analyze a particular
tree.  We still assume for now that the cut is defined by
$\incomparable{v}{w}$; we consider the other case in the next
subsection.  We have already argued that we can perform the
computation for a single bough using a number of dynamic tree
operations proportional to the number of edges incident on the bough.
Doing so will change the values $\marker{w}$.  However, we can return
them to their original values by repeating the execution of ${\proc
  MinPrecut}$ on the bough, replacing each addition operation by a
subtraction.

  It follows that we can process a collection of boughs in sequence
  since each computation on a bough starts with the $\marker{w}$
  variables back at their initial values $\cutval{\descendants{w}}$.
  Now note that every edge of $G$ is incident on at most $2$ boughs.
  It follows that the total size (number of incident edges) of all
  boughs of our tree is $O(m)$.  Therefore, the time to process all
  boughs of the tree with the algorithm of the previous section is
  $O(m\log n)$.
  
  Once we have processed all boughs, we know that we have found the
  minimum cut if even one of the two tree-edges it cuts is in a bough.
  If not, we can contract each bough of the tree (that is, merge all
  vertices in a bough into the vertex from which that bough descends)
  without affecting the minimum cut.  This can be done in $O(m)$ time
  (using a bucket sort).  It leaves us with a new tree that
  $2$-constrains the minimum cut of the contracted graph.
  Furthermore, every leaf of the new tree had at least two descendants
  in the original tree (else it would have been in a bough).  This
  means that the new tree has at most half as many leaves as the old.  It follows
  that after $O(\log n)$ iterations of the procedure for boughs
  followed by contractions of boughs, we will have reduced the number
  of tree-leaves to $0$, meaning that we will have contracted the entire
  graph.  Clearly at some time before this, we found the minimum cut.

\subsection{Comparable $v$ and $w$}

It remains to explain the procedure for finding $\precut{v}{w}$ when
(without loss of generality) $v \in\descendants{w}$.  We proceed much
as before.  In Section~\ref{sec:tree 2 slow}, we showed that 
\[
\cutval{\descendants{w}-\descendants{v}}
=
\cutval{\descendants{w}}-\cutval{\descendants{v}}
+2(\treefix{g_w}(v)-2\bothbelow{v}).
\]
It is therefore sufficient to compute, for each $v$, after factoring
out $\cutval{\descendants{v}}+4\bothbelow{v}$, the quantity
\[
\precut{v}{w}=\min_{w \in \ancestors{v}}
\cutval{\descendants{w}}+2\treefix{g_w}(v). 
\]
We again do so using the dynamic tree data structure.  We initialize
$\marker{w}=\cutval{\descendants{w}}$.  We then use dynamic tree
operations to add $2\treefix{g_w}(v)$ to $\marker{w}$. Suppose first
that $v$ is a leaf.  Then for each neighbor $u$ of $v$, we call
$\addpath{u}{2\edgeweight{v}{u}}$.  Having done so, we can extract our
answer as $\minpath{v}$.  This follows because
$\treefix{g_w}(v)=\sum_{x \in\descendants{v}}
\crossweight{x}{\descendants{w}}$.  So to add $2\treefix{g_w}(v)$ to
$w$, we want to add twice the capacity of every edge from $v$ to each
descendant $u \in \descendants{w}$.

Also as before, we can process an entire bough by working up from the
leaf.  For each vertex $v'$ in the bough, we execute
$\addpath{u'}{\edgeweight{v'}{u'}}$ operations for each neighbor $u'$
of $v'$ and then compute $\minpath{v'}$.  By the time we call
$\minpath{v'}$, we will have $\marker{w}$ properly updated for each
ancestor $w$ of $v'$.  After processing each of the boughs, we
contract all of them and recurse on the resulting tree.  Note that
unlike the case of $\incomparable{w}{v}$, here we perform only one
$\minpath{v'}$ operation for each vertex $v'$ in the bough, rather
than performing one for each of its neighbors.

\section{A Simple Class of Graphs}
\label{sec:fat}

In this section, we describe a class of graphs that are easier
to solve.  On these graphs we can find the minimum cut using only our
simple algorithm for cuts that 1-respect a tree.  Since we can
recognize these graphs quickly, in practice we might often be able to
avoid running the more complicated algorithm for the 2-respecting
case.

The Nash-Williams lower bound of $c/2$ trees in a graph with minimum
cut $c$ is often pessimistic.  The number of trees in the maximum
packing may be as large as $c$ (the exact number is determined by the
value of the ``sparsest cut'' of the graph as discussed in
Section~\ref{sec:Nash-Williams}---see~\cite{Nash:Trees} for details).
Call $G$ a {\em $\delta$-fat} graph if its maximum packing contains
more than $(1+\delta)c/2$ trees.

\begin{theorem}
  \label{thm:fat}
  For any constant $\delta>0$, the minimum cut in a $\delta$-fat
  graph can be found with constant probability in $O(m+n\log^3 n)$
  time and thus with high probability in $O(m\log n+n\log^3 n)$ time.
\end{theorem}
\begin{proof}
  Consider a maximum packing in a $\delta$-fat graph.  It has
  $(1+\delta)c/2$ trees (by weight) sharing $c$ minimum cut edges.
  Thus the average number of minimum cut edges per tree is at most
  $2/(1+\delta)<2$.  Thus, by Markov's inequality, there is a constant
  probability that the number of min-cut edges in a random tree is
  less than 2, meaning that it must be exactly 1.
  
  As before, we cannot actually take the time to find a maximum
  packing.  However, we can use the same skeleton construction as
  \ref{prop:some trees}. We have shown~\cite{Karger:Matroid} that if a
  graph is $\delta$-fat then, with high probability, any skeleton we
  construct from it is $(\delta-\epsilon)$-fat.  Consider the skeleton
  $H$ with minimum cut $c'=O(\log n)$.  Since it is $\delta$-fat, the
  maximum packing has $(1+\delta-\epsilon)c'/2$ trees.  We can
  therefore find a packing of $(1+\delta/2-\epsilon)c'/2$ trees in
  $O(n\log^3 n)$ time using the Plotkin-Shmoys-Tardos algorithm (note
  that Gabow's algorithm cannot be applied here as it always gives a
  packing of weight exactly $c'/2$ regardless of the graph's fatness).
  These trees share the at most $(1+\epsilon)c'$ edges of the cut in
  $H$ corresponding to the minimum cut of $G$.  Thus, so long as
  $\epsilon \ll \delta$ ($\epsilon = \delta/5$ suffices), the average
  number of cut edges per tree is a constant less than 2.  Thus, as in
  \ref{prop:packing}, a constant fraction of the trees (by weight)
  will 1-respect the minimum cut of $G$.
  
  Given such a tree, we can use the algorithm of Section~\ref{sec:tree
    1 slow} to identify the minimum cut in linear time.  The
  constant-probability argument follows by selecting a random tree and
  running on it our linear-time algorithm for finding the minimum cut
  that $1$-respects it.  The high probability bound follows from
  selecting $O(\log n)$ random trees and running our algorithm on
  each.  On sparse graphs, the bottleneck is now the packing algorithm
  of~\cite{Plotkin:Packing}.
\end{proof}

The above algorithm can be applied even if we do not know initially
that our graph is $\epsilon$-fat, for the heuristic is actually ``self
certifying.''  Once we have used the tree to find the minimum cut in
the graph, we can compare the size of our tree-packing to the value
$c''$ of the corresponding cut in the skeleton (which we know is
near-minimum with high probability).  If the packing has
$(1+\epsilon)c''/2$ trees, then we know the skeleton is
$\epsilon$-fat.  By a Theorem of~\cite{Karger:Matroid}, with high
probability the skeleton is $\epsilon$-fat if and only if the original
graph is.  Unfortunately, the certification is correct with high
probability but not with certainty (there is a small chance the
skeleton is $\epsilon$-fat but the graph is not).  Thus, the heuristic
does not guarantee that we have found the minimum cut.  Rather, we
have the following:

\begin{lemma}
  If a graph is $\epsilon$-fat, then with high probability the
  $\epsilon$-fat heuristic certifies itself and returns the minimum
  cut.  If a graph is not $\epsilon$-fat, then with high probability
  that $\epsilon$-fat heuristic does not certify itself.
\end{lemma}

Thus, so long as we are satisfied with a Monte Carlo algorithm, we can
halt if the algorithm certifies itself.

Although fat graphs seem to form a natural class, it should be noted
that they form a small class: a random graph (in the standard model of
Bollobas~\cite{Bollobas}) is highly unlikely to be fat.  Its minimum
cut is within $1+o(1)$ of the average degree $k$, and counting edges
shows that a graph with average degree $k$ can have at most $k/2$
spanning trees.

\section{Small Refinements}

In this section, we give two very small refinements of our algorithm
that improve its running time to the unhappy bound of $O(m(\log^2
n)\log(n^2/m)/\log\log n)$.  Such a messy bound suggests to us that
further improvements should be possible.

\subsection{Fewer trees}

Our first improvement reduces the number of trees that we must analyze
for 2-constrained cuts.

\begin{lemma}
A minimum cut can be found with high probability in $O(m\log^3
n/\log\log n+n\log^6 n)$ time.
\end{lemma}
\begin{proof}
  We use the fact that trees can be analyzed more quickly for
  1-respecting cuts than for 2-respecting cuts.  In our initial
  sampling step (discussed in Section~\ref{sec:packing}), set
  $\epsilon=1/4\log n$.  Using Gabow's algorithm to pack trees in this
  sample takes $O(n\log^7 n)$ time.  In~\cite{Karger:Skeleton} we give
  a speedup of Gabow's algorithm that improves the running time to
  $O(n\log^6 n)$.  
  
  In a second phase, choose $4 \log^2 n$ trees at random from the
  packing and find the minimum cut that 1-respects each in $O(m\log^2
  n)$ time.  In a third phase, choose $\log n/\log\log n$ trees at
  random from the packing and find the minimum cut that 2-respects
  each of them in $O(m\log^3 n/\log\log n)$ time.  Our speedup arises
  from checking $\log n/\log\log n$ trees for the ``2-respects'' case
  instead of $\log n$.  We claim that this algorithm finds the minimum
  cut with probability $1-O(1/n)$.
  
  To see this, consider the sampled graph.  It has minimum cut $c'$.
  The minimum cut of $G$ corresponds to a cut of value at most
  $(1+\epsilon)c'$ in the sample.  The tree packing has value $\rho
  \ge c'/2$, of which say $\alpha \rho$ trees $1$-respect the minimum
  cut and $\beta \rho$ trees 2-respect the minimum cut.  Note that
  since any tree which is not counted by $\alpha$ or $\beta$ must have
  at least 3 cut edges, we have
  \begin{eqnarray*}
  \alpha \rho + 2\beta \rho + 3(1-\alpha-\beta) \rho &<
  &(1+\epsilon)c' \le 2(1+\epsilon)\rho\\
  \alpha + 2\beta + 3(1-\alpha-\beta) &< &2(1+\epsilon)\\
  3-2\alpha-\beta &< &2+2\epsilon\\
  \beta &\ge &1-2\epsilon - 2\alpha\\
  \end{eqnarray*}
  
  Suppose first that $\alpha > 1/4\log n$.  Then each time we choose a
  random tree, the probability is $\alpha$ that we will choose one
  that 1-respects the minimum cut.  Thus if we choose $4\log^2 n$
  random trees, the probability that none of them $1$-respects the cut
  is at most
  \[
  (1-1/4\log n)^{4\log^2 n} < 1/n.
  \]
  Thus, with high probability in the second phase we choose and
  analyze a tree that 1-respects the minimum cut, and therefore find
  the minimum cut.
  
  Now suppose that $\alpha \le 1/4\log n$.  It follows from the
  previous inequalities and the fact that $\epsilon=1/4\log n$ that
  \begin{eqnarray*}
    \beta  &\ge &1-1/\log n\\
  \end{eqnarray*}
  It follows that when we choose $\log n/\log\log n$ trees and test
  them for 2-respecting cuts in the third phase, our probability of
  failing to choose a tree that 2-respects the minimum cut is only
  \begin{eqnarray*}
  (1-\beta)^{\log n/\log \log n} &\le &\left(\frac{1}{\log n}\right)^{\log
  n/\log\log n}\\
  &=&1/n
  \end{eqnarray*}

  Thus, regardless which case holds, we find the minimum cut with
  probability at least $1-1/n$.

\end{proof}

\subsection{Solving one tree faster}

Our second improvement reduces the time it takes to analyze a single
tree.  

\begin{lemma}
The smallest cut that 2-respects a given tree can be found in $O(m\log
n\log(n^2/m))$ time.
\end{lemma}
\begin{proof}
  We previously bounded the time to analyze a bough with $d$ incident
  edges by $O(d \log n)$.  We show that in an amortized sense, we can
  also bound this time by $O(n\log n)$.  We thus break the time to
  process a tree with up to $n$ leaves into two parts.  The time to
  reduce the number of leaves (and boughs) to $l$ is (as was shown in
  Section~\ref{sec:tree 2 fast}) $O(m(\log n)\log n/l)$.  The time to
  process the $l$-leaf tree is (by the amortized bound) $O(nl\log n)$.
  Overall, the time to process the tree is
  \[
  O((m\log n/l+nl)\log n).
  \]
  Choosing $l=m/n$ yields the claimed time bound.
  
  To prove the amortized time bound, suppose we process a bough with
  $d$ incident edges.  Afterward, we contract that bough to a single
  node.  Assuming we merge parallel edges during the contraction, a
  single node can only have $n$ incident edges.  Thus, $d-n$ of the
  edges incident on the bough vanish.  We charge the dynamic tree
  operation for each edge that vanishes to the vanishing edge, and
  charge the at-most $n$ dynamic tree operation for non-vanishing
  edges to the algorithm.  Over the entire course of the algorithm,
  each edge gets charged once (for a negligible total of $O(m)$
  operations) while the algorithm gets charged $O(nl)$ times.
\end{proof}

\section{Conclusion}

This paper has presented two algorithms for the minimum cut
problem. The first is quite simple and runs in $O(n^2\log n)$ time
(and on fat graphs in $O(m\log n)$ time).  The second algorithm is
relatively complicated, but holds out the possibility that the minimum
cut problem can be solved in linear time.  Several probably
unnecessary logarithmic factors remain in the running time, suggesting
the following improvements:
\begin{itemize}
\item Give a dynamic path-minimization data structure taking constant
  amortized time per operation.  This would reduce the running time by
  a $\log n$ factor.  We are not using the full power of dynamic trees
  (in particular, the tree we are operating on is static, and the
  sequence of operations is known in advance), so this might be
  possible.
\item Extend the approach for boughs directly to trees in order to
  avoid the $O(\log n)$ different phases needed for pruning boughs.
  This too would reduce the running time by a $\log n$ factor.
\item Give a deterministic linear-time algorithm for finding a tree
  that $2$-constrains the minimum cut.  This would eliminate the $\log
  n$ factor required by our randomized approach.  Note that
  randomization is used only to find the right tree; all remaining
  computation is deterministic.  Thus any $o(mn)$-time algorithm for
  finding a good tree would yield the fastest known deterministic
  algorithm for minimum cuts.
\end{itemize}

Our algorithm is Monte Carlo.  It would be nice to develop a
deterministic, or at least a Las Vegas, algorithm based upon our
ideas.  One standard approach to getting a Las Vegas algorithm would
be to develop a minimum cut ``certifier'' that would check the
correctness of our Monte Carlo algorithm.

Another question is whether the near-linear time algorithm can be
parallelized.  This would require finding a substitute for or
parallelization of the sequence of dynamic updates the algorithm
performs.

Our tree-packing approach has also led to some progress on the
enumeration of near-minimum cuts.  The gap between upper and lower
bounds is now extremely small and should be eliminated entirely.
Except for one graph on $7$ vertices~\cite{Nagamochi:CountCuts}, the
cycle appears to have the most small cuts, so it is the upper bound
which seems likely to be improved.

\section{Acknowledgments}

Thanks to Robert Tarjan for some helpful references and comments on
dynamic merging.  Thanks to Eric Lehman and Matt Levine for some
careful reading and suggestions for presentation improvements.

\appendix

\section{Optimality of the Cut Counting Proof}

In Theorem~\ref{thm:countcuts}, we gave an upper bound of 
\[
q_k = \frac{\sum_{r \le k}(k+1-r)\binom{n}{r}}{k+1-2\alpha}
\]
on the number of $\alpha$-minimum cuts a graph may have.  The upper
bound held for every (and thus the smallest) value of
$k\ge\floor{2\alpha}$ (to ensure $q_k>0$).  The proof of this bound
relied on a weighted bipartite graph on cuts and packed trees.  It is
not obvious that this proof has been ``optimized.''  Perhaps a
different assignment of edge weights would yield a better bound.  In
this section, we show that no better bound is possible.  Our analysis
was derived using linear-programming duality, but is presented in a
simplified way that does not use duality.

Our upper-bound proof can be thought of as a kind of game.  We choose
to assign a weight $w_r \ge 0$ to any edge connecting a cut $C$ to a
packed tree containing exactly $r$ edges of $C$.  The maximum weighted
tree degree is therefore $\sum w_r\binom{n}{r}$.  We then consider the
smallest possible weighted degree that a cut $C$ could have given
these weights.  If $n_r$ trees contain exactly $r$ edges of this
minimum-degree cut, then its weighted degree is $\sum w_r n_r$.
We deduce that the number of cuts is upper bounded by
\[
R(\set{w_r,n_r})=\frac{\frac{c}{2}\sum w_r \binom{n}{ r}}{\sum w_r
  n_r}.
\]
The bound only holds if $n_r$ defines the {\em smallest
  possible} cut degree, so our choice of weights $w_r$ yields an upper
  bound of
\[
\max_{n_r} R(\set{w_r,n_r})
\]
on the number of cuts.  Since we are interested in as tight a bound as
possible, we would actually like to compute
\[
\min_{w_r}\max_{n_r} R(\set{w_r,n_r}).
\]
Of course, we need not consider all possible values $n_r$.  In
particular, since every tree has $r$ cut edges for {\em some} $r$, we
know that
\begin{align}
\sum_r n_r &= c/2. \label{eq:count trees}\\
\intertext{Furthermore, since the trees must share the at most $\alpha
  c$ cut edges, we must have}
\sum_r r\cdot n_r &\le \alpha c. \label{eq:count edges}\\
\intertext{Finally, we must clearly have}
n_r \ge 0 \label{eq:positive}
\end{align}
for every $r$.  Conversely, for any $n_r$ satisfying these three equations,
there clearly exist trees yielding these values $n_r$.

Using these restrictions on $n_r$, we used a particular set of values
$w_r$ to show in Theorem~\ref{thm:countcuts} that
\[
\min_{w_r}\max_{n_r} R(\set{w_r,n_r}) \le \min_{k\ge\floor{2\alpha}} q_k.
\]
We now prove that one cannot make a stronger argument.
We exhibit a particular assignment of values $n_r$ for which 
\[
\min_{w_r} R(\set{w_r,n_r}) = \min_{k\ge\floor{2\alpha}} q_k.
\]
Thus, no weight assignment can prove a ratio better than $\min q_k$.
Since we have given weights that prove a ratio of $\min q_k$, we have
an optimal bound.

To show our analysis is optimal, fix $k$ to minimize $q_k$ and
consider the following values for $n_r$:
\begin{align*}
n_r&=\frac{c}{2q_k}\binom{n}{r} \qquad (r=1,\ldots,k)\\
n_{k+1} &= \frac{c}{2}\left(1- \frac1{q_k}\sum_{r \le k}\binom{n}{r}\right)\\
n_r &= 0 \qquad \qquad (r > k+1)
\end{align*}
We will shortly show that these $n_r$ satisfy the three feasibility
equations and thus correspond to some set of trees.  We will also
prove one more fact: for every $r$, 
\begin{align}
n_r \le \frac{c}{2q_k}\binom{n}{r}. \label{eq:ratio}
\end{align}
This clearly holds with equality for $r < k+1$ and is trivial for
$r>k+1$; we will show it holds for $r=k+1$ as well.  Assuming it for
now, observe that for any weights $w_r$ used in the proof, the ratio
we get for our proof is
\begin{align*}
R(w_r,n_r)&=\frac{\frac{c}{2}\sum \binom{n}{r}w_r}{\sum n_r w_r}\\
&\ge\frac{\frac{c}{2} \sum_{r \le k}\binom{n}{r} }{\sum_{r \le k}
  \frac{c}{2q_k}\binom{n}{r}}\\
&= q_k.
\end{align*}
In other words, no weight assignment can yield a better bound than $q_k$
in the proof of Theorem~\ref{thm:countcuts}.

We now need to justify the claims made earlier in
Equations~\ref{eq:count trees},~\ref{eq:count
  edges},~\ref{eq:positive}, and~\ref{eq:ratio}.
Equation~\ref{eq:count trees} is obvious from the definition of
$n_{k+1}$.  Next we prove Equation~\ref{eq:count edges}, showing that
$\sum r\cdot n_r \le \alpha c$.  Simply note
\begin{align*}
\sum r\cdot n_r &= \sum_{r \le k} r \frac{c}{2q_k} \binom{n}{r}+
(k+1)\frac{c}{2}(1- \frac1{q_k}\sum_{r \le k}\binom{n}{r})\\
&=\frac{c}{2}\left(\sum_{r \le k} \frac{r}{q_k} \binom{n}{r}+k+1-(k+1)\sum_{r \le
  k}\frac1{q_k} \binom{n}{r}\right)\\
&=\frac{c}{2}\left(k+1-\sum \frac{k+1-r}{q_k}\binom{n}{r}\right)\\
&=\frac{c}{2}(k+1-(k+1-2\alpha))\\
&=\alpha c
\end{align*}
as claimed.

It remains only to prove  Equations~\ref{eq:positive}
and~\ref{eq:ratio}, which are non-obvious only for $r=k+1$.  Plugging
in the definition of $n_{k+1}$, we need to prove that
\[
0 \le \frac{c}{2}(1- \frac1{q_k}\sum_{r \le k}\binom{n}{r})
\le\frac{c}{2q_k}\binom{n}{k+1}
\]
or equivalently (multiplying by $2q_k/c$) that
\[
\sum_{r \le k}\binom{n}{k} \le q_k \le \sum_{r \le k+1} \binom{n}{r}.
\]
We begin with the upper bound. Since $q_k \le q_{k+1}$ (by choice of
$k$ to minimize $q_k$), we deduce that
\begin{align*}
q_k &= (k+2-2\alpha)q_k-(k+1-2\alpha)q_k\\
&\le (k+2-2\alpha)q_{k+1}-(k+1-2\alpha)q_k\\
&=\sum_{r \le k+1}(k+2-r)\binom{n}{r}-\sum_{r \le
  k}(k+1-r)\binom{n}{r}\\
&=\sum_{r \le k+1}\binom{n}{r}
\end{align*}
as desired.  Now consider the lower bound.  If $k>\floor{2\alpha}$,
then we must have $q_k \ge q_{k-1}$, so we can work much as before:
\begin{align*}
q_k &= (k+1-2\alpha)q_k-(k-2\alpha)q_k\\
&\ge (k+1-2\alpha)q_{k+1}-(k+2\alpha)q_k\\
&=\sum_{r \le k}(k+1-r)\binom{n}{r}-\sum_{r \le
  k-1}(k-r)\binom{n}{r}\\
&=\sum_{r \le k}\binom{n}{r}
\end{align*}
If, on the other hand $k=\floor{2\alpha}$, then just recall that
\[
q_k= \sum_{r \le k}\frac{k+1-r}{k+1-2\alpha}\binom{n}{r}.
\]
But note that for all $r\le k$, the numerator $k+1-r \ge 1$ while the
denominator $k+1-2\alpha \le 1$.  So the claim holds.

This completes the analysis of our proof, and shows that now other
weight function assignment will improve on the one we have already
given.  Of course, this does not prove that there is no better bound
on the number of cuts: a completely different proof could show one.

\bibliographystyle{alpha} \bibliography{me}

\end{document}